# Direct Evidence Delay with A Task Decreases Working Memory Content in Free Recall

Eugen Tarnow, Ph.D.[1]

18-11 Radburn Road, Fair Lawn, NJ 07410, USA

etarnow@avabiz.com

[1] The author is an independent researcher. He received a PhD in physics from MIT. His research areas include semiconductor physics, marketing science, management science, and the psychology of dreams, conformity, obedience and short term memory.

# Abstract

Recently it was shown that free recall consists of two stages: the first few recalls empty the working memory and a second stage concludes the recall (Tarnow, 2015; for a review of the theoretical prediction see Murdock, 1974). It is commonly believed that a delay with a task before the recall starts removes the content of working memory (Glanzer & Cunitz, 1966). Here is presented the first direct evidence that this is indeed the case.

# Introduction

Until recently free recall stood out as one of the great unsolved mysteries of modern psychology (Hintzman; 2011, for reviews, please see, for example, Watkins, 1974; Murdock, 1974; Laming, 2010). Items in a list are displayed or read to subjects who are then asked to retrieve the items. It is one of the simplest ways to probe short term memory. The results (Murdock, 1960; Murdock, 1962; Murdock, 1974) have defied explanation. Why do we remember primarily items in the beginning and in the end of the list, but not items in the middle, creating the famous u-shaped curve of probability of recall versus serial position? Why can we remember 50-100 items in cued recall but only 6-8 items in free recall?

Some of the mystery has been removed. We now know explicitly that free recall consists of two stages (Tarnow, 2015; for a review of the experiments and theory which predicted the two stages see Murdock, 1974). In the first stage working memory is emptied and in the second stage a different retrieval process occurs. Working memory is responsible for the recency part of the serial position curve and for some of the first item recall when using short lists (Tarnow, 2015).

In order to isolate the (still mysterious) second stage, there is a common variation on a free recall experiment called “delayed free recall”. Glanzer and Cunitz (1966) invented the delay manipulation to test the hypothesis that free recall has “two storage mechanisms”. They found out that pure delay was not as effective as a delay with a task: their delay task consisted of counting out loud starting with a random single digit. After 10 seconds most of the recency peak was gone and after 30 seconds nothing remained of the recency peak.

In this contribution I will show explicitly that a delay with a task removes almost all working memory content. I will also show that though working memory content is removed, some of the items that would have been in working memory can be retrieved in the second stage, an overlap that was predicted earlier (reviewed by Murdock, 1974).

# Method

This article makes use of the Howard & Kahana (1999) data set (downloaded from the Computational Memory Lab at the University of Pennsylvania (http://memory.psych.upenn.edu/DataArchive). In Table 1 is summarized the experimental processes which generated the data set.

| *Work* | *Item types* | *List length* | *Presentation interval* | *Interval between last presented item and recall* | *Recall interval* | *Item presentation mode* | *Procedure (quoted from data source)* |
|---|---|---|---|---|---|---|---|
| Howard and Kahana (1999) experiment 1 | Toronto Noun Pool | 16 | 1 item per second | 16 second delay or no delay | 45 seconds | Visual | “During list presentation, participants were required to perform a semantic orienting task on the presented words. The participants were to press the left control key if they judged the word to be concrete and the right control key if they judged it to be abstract. The presentation rate of the items was not dependent on the concreteness judgments. In the immediate condition, participants were cued to begin recall immediately after list presentation. Recall was cued with the presentation of three asterisks accompanied by a 500-ms tone. Participants were given 45 s to recall as many items as possible from the list. Vocal responses were recorded for later scoring … In the delayed condition, before free recall, participants were given an arithmetic distractor task that lasted at least 10 s. In this task, participants made true–false judgments on simple arithmetic equations as quickly and as accurately as possible.” |

Table 1. Properties of the Howard & Kahana (1999) dataset.

# Results

In Fig.1 is displayed the serial position curve for immediate recall, recall by recall. By definition, the first recall is working memory and from Fig. 1 it seems that more than 2 but less than 4 items are retrieved from working memory. In Fig. 2 is displayed the serial position curve for the delayed recall, recall by recall. In this case more than 0 but less than 1 item is retrieved from working memory, a difference of about 2 items.

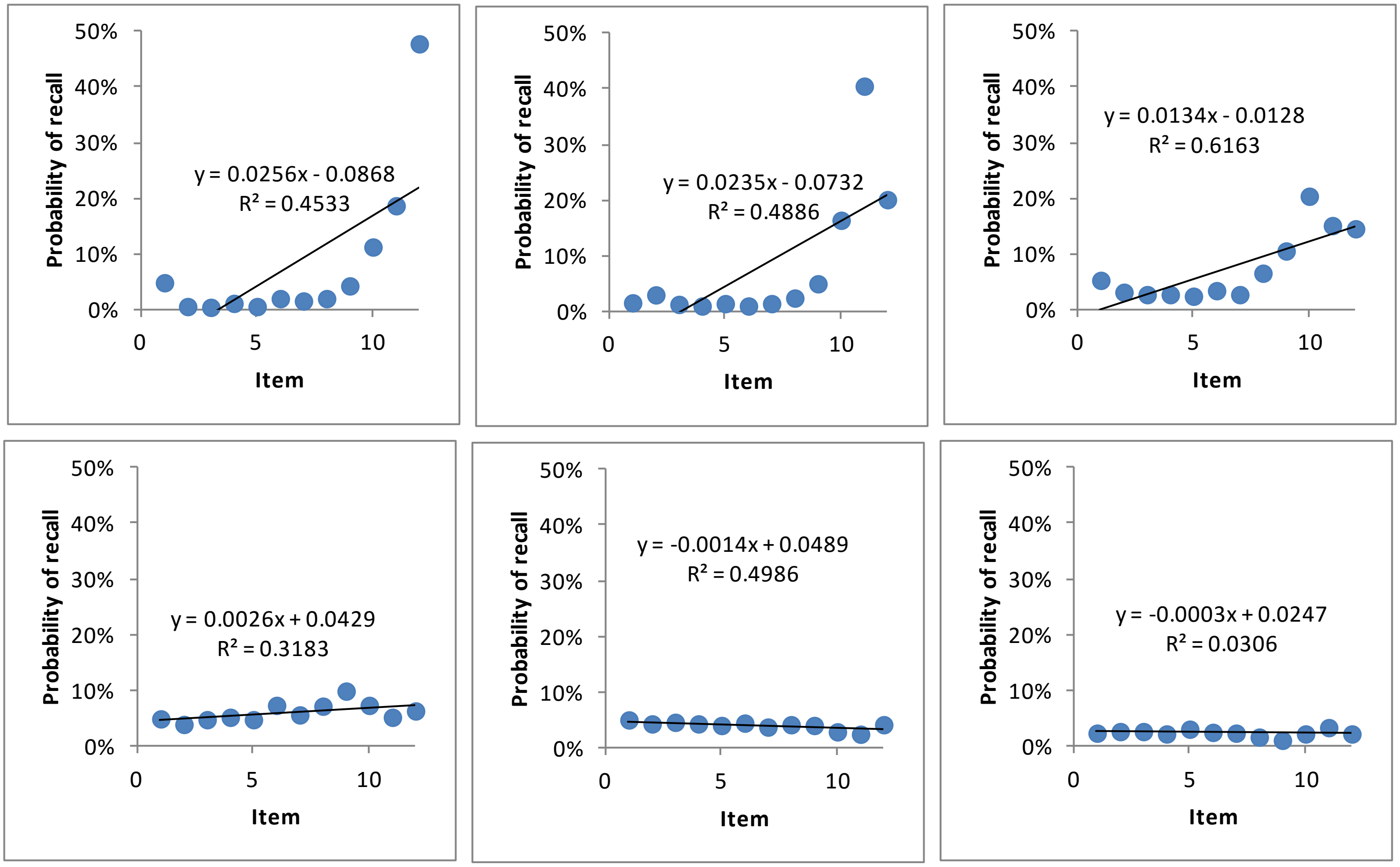

*Fig. 1. Immediate recall: top three panels correspond to recalls 1-3 (left to right) and the bottom three panels correspond to recalls 4-6 (left to right).*

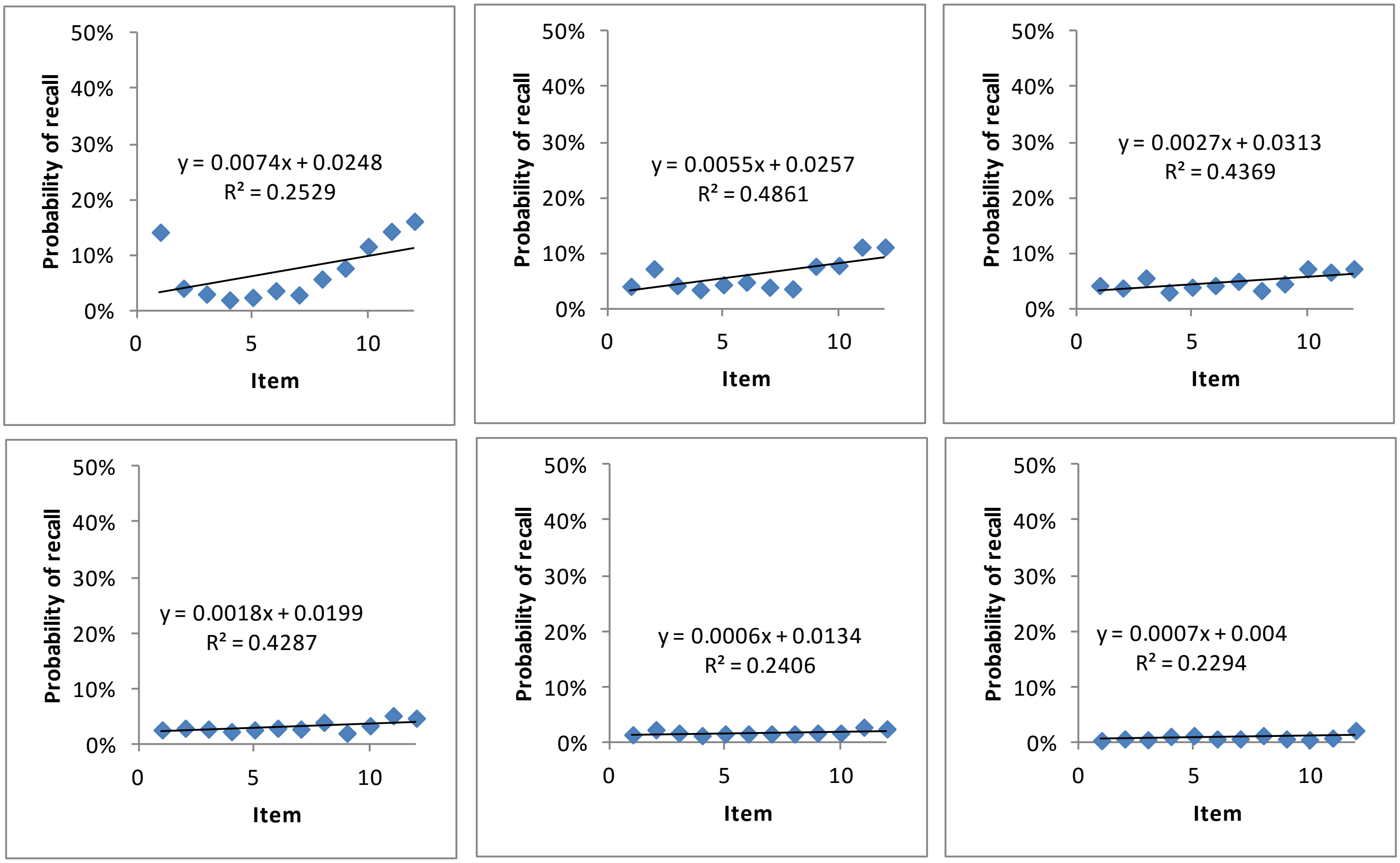

*Fig. 2. Delayed recall: top three panels correspond to recalls 1-3 (left to right) and the bottom three panels correspond to recalls 4-6 (left to right).*

In Fig. 3 is plotted the fitted slopes to the individual recalls. The cross-over from working memory to the second stage is 3 items for the immediate free recall and less than one item for delayed free recall. This yields a second calculation of the difference in working memory content of about 2.5 items.

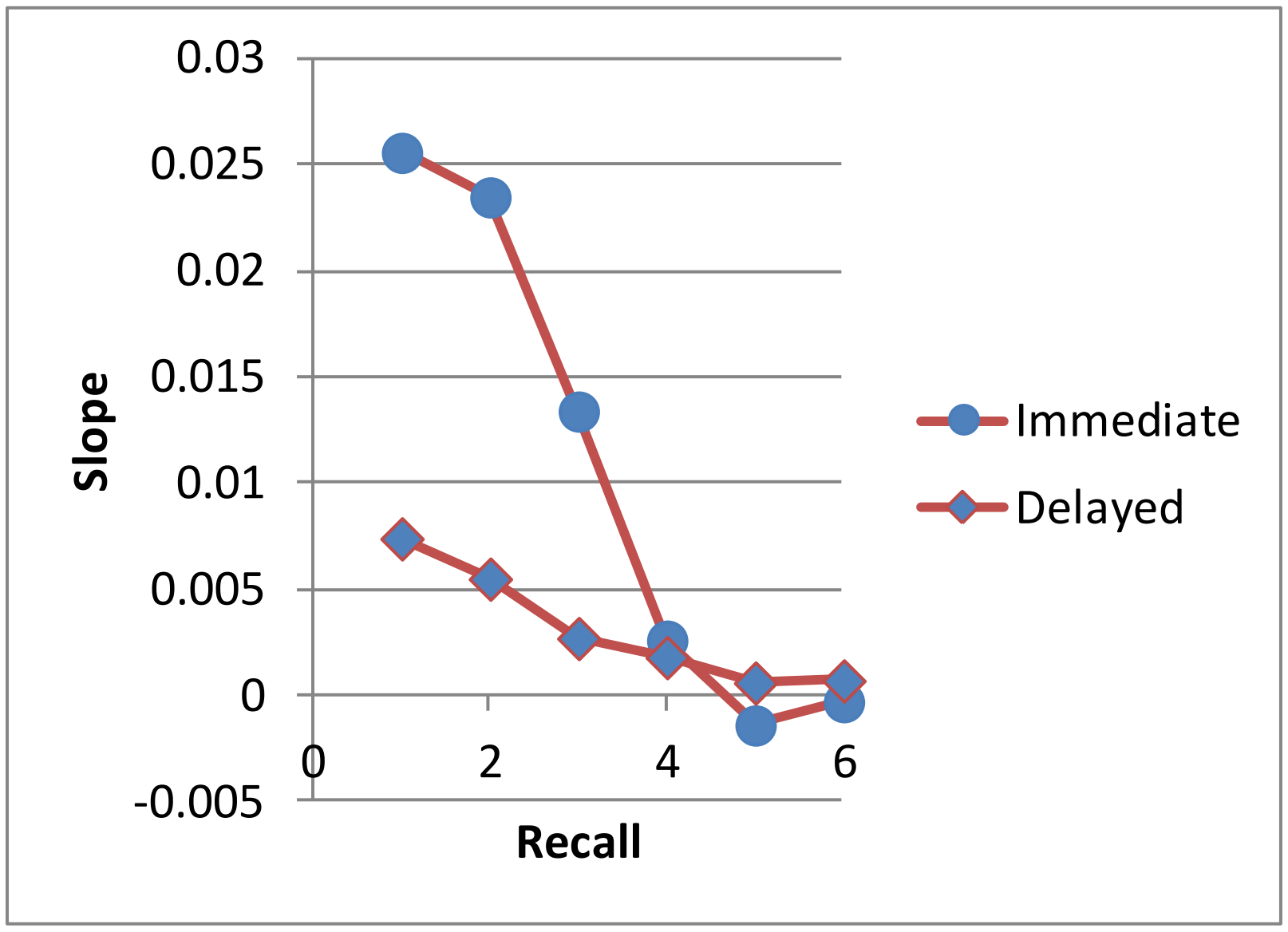

*Fig. 3. Slope of linear fit as a function of recall. Note the similarity of the immediate curve to a rounded step function with a half point at the third recall.*

In Fig. 4 is shown the total recall for immediate recall, delayed recall (which shows lessened recency) and the difference between the two.

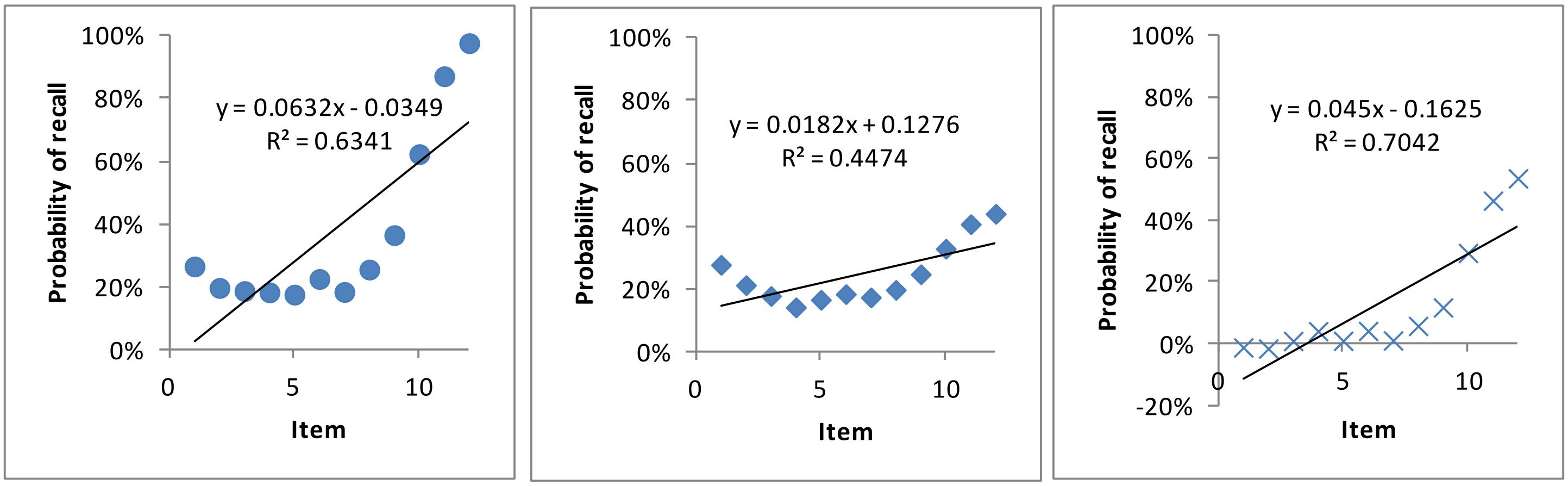

*Fig. 4. Total recall for immediate free recall (left panel), for delayed free recall (middle panel) and the difference between immediate and delayed free recall (right panel). Note the total recall difference between immediate and delayed conditions is 1.6 words, much smaller than working memory.*

# Discussion

As had been theorized and shown indirectly before (Glanzer and Cunitz, 1966; for a review, see Murdock, 1974), free recall delayed with a task removes some working memory items from the recall. The amount removed from working memory in the Howard & Kahana (1999) experiment was calculated to be 2-2.5 items.

The overall recall decreased only by 1.6 items. The difference between the smaller decrease of total recall and the larger decrease in the number of working memory items presumably shows that some of the items that had been in working memory were also accessible in the second stage of the recall. This has previously been predicted (see Murdock, 1974, for a review). In addition, the signal to recall may function as a cue to the first item (note the high probability of first item recall during the first recall in the delay condition in Fig. 2) and this cue may be ignored if working memory is full.

Individual recall distributions as presented here should be useful for future delayed recall experiments to show how effective a particular delay technique is in removing content from working memory.